\pdfoutput=1
\documentclass{article}
\usepackage{arxiv}

\usepackage[utf8]{inputenc} 
\usepackage[T1]{fontenc}    
\usepackage{hyperref}       
\usepackage{url}            
\usepackage{booktabs}       
\usepackage{amsfonts}       
\usepackage{nicefrac}       
\usepackage{microtype}      
\usepackage{lipsum}
\usepackage{graphicx}
\usepackage{enumitem}
\usepackage{siunitx}
\usepackage{bm}
\usepackage{float}
\DeclareUnicodeCharacter{2212}{-}
\usepackage{amsmath}
\usepackage{caption}
\usepackage{subcaption}
\usepackage[numbers]{natbib}

\title{Polarization and reliability of news sources in Wikipedia}

\author{
 Puyu Yang \\
  Institute for Logic, Language and Computation (ILLC)\\
  University of Amsterdam\\
  LAB42, 1098 XH Amsterdam \\
  \texttt{p.yang2@uva.nl} \\
   \And
 Giovanni Colavizza \\
  Institute for Logic, Language and Computation (ILLC)\\
  University of Amsterdam\\
  LAB42, 1098 XH Amsterdam \\
  \texttt{g.colavizza@uva.nl}\\
}

\date{}

\begin{document}
\maketitle
\begin{abstract}
Wikipedia is the largest online encyclopedia: its open contribution policy allows everyone to edit and share their knowledge. A challenge of radical openness is that it facilitates introducing biased contents or perspectives in Wikipedia. Wikipedia relies on numerous external sources such as journal articles, books, news media, and more. News media sources, in particular, take up nearly third of all citations from Wikipedia. However, despite their importance for providing up-to-date and factual contents, there is still a limited understanding on which news media sources are cited from Wikipedia. Relying on a large-scale open dataset of nearly 30M citations from English Wikipedia, we find a moderate yet systematic liberal polarization in the selection of news media sources. We also show that this effect is not mitigated by controlling for news media factual reliability. Our results contribute to Wikipedia's knowledge integrity agenda in suggesting that a systematic effort would help to better map potential biases in Wikipedia and find means to strengthen its neutral point of view policy.
\end{abstract}


\section{Introduction}
Wikipedia is one of the most extensive encyclopedias worldwide, providing an open go-to reference for reliable online content and a key hub to the Web~\citep{piccardi2020quantifying}. Wikipedia's articles are contributed by volunteers, following the policies of taking a neutral point of view (NPOV), verifiability of facts and sources, and contributing no original research. In principle, all Wikipedia articles should be ``based on reliable, independent, published sources with a reputation for fact-checking and accuracy''.\footnote{\url{https://en.wikipedia.org/wiki/Wikipedia:Verifiability##What_counts_as_a_reliable_source}.} Sources are usually cited in footnotes and references. News media outlets provide a sizeable share of Wikipedia's cited sources, yet they often contain both factual contents and opinions or viewpoints around them~\citep{fetahu2015much}. News reporting from well-established outlets is generally considered reliable for statements of fact. However, a potential for viewpoint bias remains and may affect the integrity of knowledge in Wikipedia, or at least Wikipedia's neutral point of view. While millions of volunteer contributors create and maintain free knowledge in Wikipedia~\citep{aragon2021preliminary}, new challenges emerge in terms of information quality and reliability~\citep{saez2019online,morgan2019research}. What is more, news media often polarize around opposite political viewpoints~\citep{patterson2011out,sutter2000can} and may be exposed to misinformation or fake news~\citep{Lazer_fake}. While previous work has focused on assessing the reliability of Wikipedia contents~\citep{przybylacountering} and editors' possible biases~\citep{rogers2012neutral,yasseri2014most}, researchers still have to systematically investigate Wikipedia's knowledge integrity~\citep{wikipedia.org,sugandhika2022assessing}. We contribute to this line of work by exploring the political polarization of news media sources in Wikipedia. We further assess their reliability and determine whether there is a relationship between the two effects. To this end, we ask the following research questions:

\begin{enumerate}
    \item RQ1: Is there political polarization in the news media sources cited in Wikipedia?
    \item RQ2: What factors influence news media polarization in Wikipedia? Specifically, is there a relationship between news media political polarization and factual reliability?
\end{enumerate}

In order to answer these questions, we rely on the large-scale dataset \textit{Wikipedia Citations}~\citep{singh2021wikipedia}; we use third-party sources to estimate the political polarization and reliability of news media outlets: the Media Bias Monitor (MBM) and the Media Bias Fact Check (MBCF). Following the approach taken by MBM, we consider political polarization across a mono-dimensional spectrum between liberal left and conservative right, acknowledging this as a limitation. Reliability conveys an estimate of the factual correctness of an outlet, in terms of contents and framing. We speculate that reliability and polarization might be related, for example with media outlets closer to a given political leaning being considered more reliable on average. Firstly, we provide a quantitative overview of news media sources' political polarization in Wikipedia (RQ1); secondly, we make use of regression analysis to clarify the relationship between news media source polarization, on the one hand, and an article's topic, WikiProject and a source factual reliability, on the other hand (RQ2). Our aim is to inform Wikipedia's knowledge integrity agenda~\citep{wikipedia_Knowledge_Integrity} by rising awareness on possible biasing effects in Wikipedia's sources.

\section{Previous work}

\subsection{Wikipedia's core policies}
Wikipedia strives to take a neutral viewpoint and provide reliable contents~\citep{mesgari2015sum}. To this end, Wikipedia abides to three core content policies:
\begin{enumerate}
    \item{Neutral Point of View (NPOV): ``representing fairly, proportionately, and, as far as possible, without editorial bias, all of the significant views that have been published by reliable sources on a topic.''}\footnote{\url{https://en.wikipedia.org/wiki/Wikipedia:Neutral_point_of_view}.}
    \item{Verifiability: ``other people using the encyclopedia can check that the information comes from a reliable source.''}\footnote{\url{https://en.wikipedia.org/wiki/Wikipedia:Verifiability}.}
    \item{No original research: ``Wikipedia articles must not contain original research.''}\footnote{\url{https://en.wikipedia.org/wiki/Wikipedia:No_original_research}.}
\end{enumerate}

These three policies could help to improve Wikipedia's article quality~\citep{pavalanathan2018mind} and could enable us to collectively address many of the practical issues stemming from collaboratively curating encyclopedic content~\citep{arazy2006wisdom}. However, from an epistemic perspective, they lay the responsibility for assessing content quality and reliability to third parties  via reliable sources~\citep{saez2019online}. On the one hand, reliable sources are essential to Wikipedia's status of a neutral encyclopedia, yet on the other hand the selection of sources invariably leads to controversies~\citep{borra2014contropedia} and even edit wars~\citep{sumi2011edit}. To be sure, this might largely be a feature as some researchers believe that the existence of such controversies ultimately leads to better quality articles~\citep{shi2019wisdom}.

\subsection{Knowledge integrity in Wikipedia}
As one of the main repositories of free knowledge available today, Wikipedia plays a central role on the Web~\citep{arazy2006wisdom,smith2020situating}. Its very widespread usage and radical openness to readers and contributors make Wikipedia vulnerable to malicious information attacks and disinformation~\citep{saez2019online}, which in turn could compromise Wikipedia's knowledge integrity~\citep{aragon2021preliminary}. 

Knowledge Integrity is one of the research priorities individuated by Wikimedia Research, whose aim is to identify and address threats to contents in Wikipedia, to increase the capabilities of patrollers, and to provide mechanisms for assessing the reliability of sources~\citep{wikipedia_Knowledge_Integrity}. As of August 2022, Wikipedia is active in 318 language versions and each version is maintained by a dedicated (or language-specific) community~\footnote{\url{https://en.wikipedia.org/wiki/List_of_Wikipedias##cite_note-3}.}, thus, knowledge integrity risks can arise in many different forms. If we compare the size of the active editor communities with the scale of the Wikipedia project, it is clear that resources for patrolling and verifying contents remain on high demand~\citep{morgan2019research,saez2019online}. Besides, a lack of geographical diversity might favor nationalistic biases~\citep{Non-English}. From the perspective of contents, disputes between community members due to disagreements about the content of articles~\citep{rogers2012neutral,yasseri2014most}, content verifiability~\citep{lewoniewski2019multilingual,redi2019citation} and quality~\citep{lewoniewski2017relative,rogers2012neutral} are significant and enduring aspects of Wikipedia.

\subsection{Wikipedia's sources}
The `verifiability' policy guarantees the existence of an important aspect of Wikipedia: citations~\citep{kaffee2021references}. Citations serve several important roles: ``they uphold intellectual honesty and reduce the risk of plagiarism, they attribute prior work and ideas to their authors, they allow the reader to  independently determine whether the referenced material supports the statements made by an editor in Wikipedia, and thus they help the reader gauge the strength and validity of the material an editor has relied on.''\footnote{\url{https://en.wikipedia.org/wiki/Citation}.} However, evidence shows that references in Wikipedia are not too actively used by readers~\citep{piccardi2020quantifying}. In Wikipedia, scientific or scholarly literature takes up a large proportion of citations to sources~\citep{nielsen2017scholia,singh2021wikipedia}, and Wikipedia's citation rates are often aligned with those in the scholarly literature~\citep{shuai2013comparative,mesgari2015sum,yang2021map}. Although it has been found that Wikipedia can have an influence on scientific research~\citep{thompson2018science}, and some professional journalists have also begun to use Wikipedia in their work~\citep{messner2011legitimizing}, the debate on using Wikipedia as a credible academic information resource is still active~\citep{tomaszewski2016study}. 

Despite the efforts of Wikipedia's contributors, many or even most articles in Wikipedia may still contain unsubstantiated or outdated claims, especially those flagged as being of lower quality~\citep{lewoniewski2017relative}. Sometimes editors' might not use citations systematically~\citep{chen2012citation,forte2018information} or engage in polarized edit conflicts~\citep{Umarova}. Research suggests that some editors' violations might be caused by biases~\citep{hube2017bias}, such as cultural~\citep{callahan2011cultural}, political~\citep{greenstein2012collective} or gender bias~\citep{wagner2015s}.  

Das and Lavoie~\citep{das2014automated} determine the topics an editor is interested in and the editor’s stance by editors' behaviour and interactions, finding that bias exists especially when a single point of view dominates controversial topics. Hube~\citep{hube2017bias} provides a method to detect both explicit and implicit bias in Wikipedia articles and observe its evolution by analyzing language, editing and citation styles. Greenstein and Zhu~\citep{greenstein2012collective} analyse political bias in Wikipedia by measuring the degree of political leaning of an article. They rely on a content-based method~\citep{gentzkow2010drives} which calculates the frequency of particular phrases to measure the degree of political bias. They find that Wikipedia had a liberal bias in the early years, but that bias declines over time, supporting ``a narrow interpretation of Linus' Law, namely, [that] a greater number of contributors to an article makes an article more neutral''. Besides, since Wikipedia has many language versions, different language versions can also contain specific biases. A study on the Wikipedia pages of UK politicians surfaced a substantial polarization of editors across political lines, in turn reflected in their choice of news media sources~\citep{Agarwal}. Ewa and Susan~\citep{callahan2011cultural} find systematic biases in the focus on a particular topic or person in Wikipedia versions in different languages. Zhou et al.~\citep{zhou2015wikipedia} find that people's attention to war-related topics affects the number of words and the number of subjective concepts, which in turn affects the bias of emotional expression. Last but not least, several studies have analyzed gender differences in Wikipedia. Wagner et al.~\citep{wagner2015s} find that while women on Wikipedia are covered and featured well in many Wikipedia language editions, the way women are portrayed might differ from the way men are portrayed. Reagle and Rhue~\citep{reagle2011gender} study gender bias by comparing Wikipedia and Encyclopedia Britannica. They illustrate that while the number of articles related to women is increasing, compared to the articles on men, the articles on women are more likely to be missing on Wikipedia. Researchers have also made efforts to improve the verifiability of Wikipedia's contents for example by flagging unsupported contents in view of adding citations to reliable sources~\citep{fetahu2016finding,redi2019citation}.

\subsection{News media sources in Wikipedia}

Wikipedia supports the use of sources from news media outlets: ``news reporting from well-established news outlets is generally considered to be reliable for statements of fact.''\footnote{\url{https://en.wikipedia.org/wiki/Wikipedia:Identifying_reliable_sources}.} News media sources are indeed among the most-used in Wikipedia. Fetahu et al.~\citep{fetahu2015much} find that almost 20\% of the external references in the English version of Wikipedia are to news articles. In the dataset we use for this contribution such proportion is closer to 30\%. Our previous work also found that Wikipedia's news sources are overall factually reliable, yet not uniformly so~\citep{yang_puyu_2022_6912664}. Nielsen uses a Wikipedia dump from 2008 to find that the BBC, the New York Times, and the Washington Post were the most cited news media outlets at the time, with the BBC far ahead of the other outlets. Among the top 20 most-cited news outlets, most are American and four each being Australian and British~\citep{Top_news}. 
However, it is difficult to ignore the potential for polarization in news media~\citep{patterson2011out,wolton2019biased}, as well as their uneven reliability across the spectrum of outlets~\citep{lazer2018science}. Previous work has focused on providing methods to assess the reliability of Wikipedia's content. Hube and Fetahu~\citep{hube2018detecting} proposed a supervised classification approach based on a self-build bias word lexicon, which could be able to detect biased statements with an accuracy of 74\%. Przybyła et al.~\citep{przybylacountering} collect a corpus of over 50 million citations to 24 million identified sources from Wikipedia Complete Citation Corpus (WCCC) and build a search index using multiple meaning representations, using NLP (Natural Language Processing) and ML (Machine Learning), enabling the automatic retrieval of sources to support or disprove a claim. While a considerable amount of work has been done to assess the polarization and reliability of Wikipedia's contents separately, their systematic and combined assessment for news media sources remains an open challenge.

\section{Data}

Our work is primarily based on \textit{Wikipedia Citations}, a public dataset of citations from English Wikipedia to all its sources, including news media~\citep{singh2021wikipedia}. We enrich \textit{Wikipedia Citations} with data from the Media Bias Monitor (MBM) and the Media Bias Fact Check (MBFC): two authoritative indices of news media outlets providing an estimate of their political leaning and factual reliability. The combination of these sources allows us to quantify the political polarization and factual reliability of news sources cited from Wikipedia.

\subsection{Wikipedia Citations}
\textit{Wikipedia Citations} includes more than 29M citations extracted from the over 6M articles English Wikipedia in May 2020. In \textit{Wikipedia Citations}, each Wikipedia page contains several citations pointing to external sources. Of these, 25M (85.2\%) are equipped with external links (URLs). For these URLs we extract the domain name using the \textit{tldextract} package.\footnote{\url{https://pypi.org/project/tldextract}.} Domain names are critical in our approach as they allow linking to the Media Bias Monitor (MBM) and Media Bias Fact Check (MBFC) indices, which are based on Web domains and sub-domains. For example, given the external link \textit{https://www.nbcnews.com/politics/politics-news/al-gore-compares-climate-deniers-uvalde-law-enforcement-officers-nobod-rcna39707}, after domain extraction we end up with \textit{www.nbcnews.com}. With this method, we extract 1,554,632 unique domains. Then, we link domain names to MBM and MBFC by querying their APIs. The main limitation of this approach is that it works at the news media outlet level, which corresponds to the domain name, and not at the specific source level (the actual cited news article). 

We further enrich \textit{Wikipedia Citations} with information about an article's topics and WikiProjects, following the procedure described in our previous work~\cite{yang2021map}. With the data from ORES Web service\footnote{\url{https://wiki-topic.toolforge.org/##lang-agnostic-model}.} and public data~\citep{Johnson2020}, we equip Wikipedia articles with topic (coverage of 99.3\%) and WikiProjects (coverage of 17.5\%). In this paper, we also use fractional counting to account for an article belonging to multiple topics or projects at the same time.

\subsection{Media Bias Monitor (MBM)}
To estimate the political polarization of Wikipedia citations, we use the Media Bias Monitor~\citep{ribeiro2018media}. This system collects demographic data about the Facebook followers of 20,448 distinct news media outlets via Facebook Graph API~\footnote{\url{developers.facebook.com/docs/graph-api}.} and Facebook Marketing API~\footnote{\url{developers.facebook.com/docs/marketing-apis}.}. These data include political leanings, gender, age, income, ethnicity and national identity. For political leanings, the Facebook Audience API~\footnote{\url{developers.facebook.com/docs/marketing-api/audiences-api}.} provides five levels: Very Conservative, Conservative, Moderate, Liberal, Very Liberal. To measure the political leaning of an outlet, MBM firstly finds the fraction of readers having different political leanings, and then multiply the fraction for each category with the following values: very liberal (−-2), liberal (-−1), moderate (0), conservative (1), and very conservative (2). The sum of such scores provides a single polarization score for the outlet, ranging between −-2 and 2, where a negative score indicates that a media outlet is read more by a liberal leaning audience, while a positive score indicates a conservative leaning audience. 

In the original paper, MBM is compared to alternative approaches used to infer the political leanings of news media outlets, finding that this method highly correlates with most alternatives. What is more, MBM covers 20,448 news media outlets and provides multi-dimension data. On these grounds, we use MBM in our study. As we have 1,554,632 unique domains from \textit{Wikipedia Citations}, which mostly are not news media outlets, we focus the matching of unique domains to MBM on the most frequently cited domains from Wikipedia. In Figure~\ref{fig:cumulative distribution of unique domains}, we show that the top 140,000 unique domains cover up to 90\% of all \textit{Wikipedia Citations}. We thus decided to only keep these top 140,000 domain names for our study, and match them in MBM. 

When looking a domain name up via the MBM API~\footnote{\url{https://twitter-app.mpi-sws.org/media-bias-monitor/index.php}.}, we can get four different query results:

\begin{enumerate}
\item No match; example: \textit{trove.nla.gov.au}.
\item One exact match; example: \textit{www.breitbart.com}.
\item More than one result, including the exact match; example: \textit{www.abc.com}.
\item More than one result, \textit{not} including an exact match; example: \textit{www.nytimes.com}.
\end{enumerate}

In each case, we proceed as follows. We label domains without a match (result 1) as \textit{NaN}, while we use the polarization score of the exact match for domains with result 2 or 3. For domains with result 4, we use the average polarization score of all the matches, under the assumption that this approximates the polarization score of the exact match. To test our assumption, we use the 1113 unique domains that have multiple results including an exact match (result 3), and compare the distribution of exact polarization scores and average polarization scores in Figure~\ref{fig:Distribution of bias score for multiple results}. We can see that the two distributions are overall comparable, which supports our assumption. Following this procedure we are able to equip 4,866,377 citations (16.6\% out of a total of 29.3M) with polarization scores. These 4.9M citations are all the citations in Figure~\ref{fig:Figure2_right}, while 29.3M are 100\% of citations in Figure~\ref{fig:Figure2_left}. We note again that 29.3M is the total number of citations in the dataset, while citations to news media sources are estimated at 8.9M~\cite{yang_puyu_2022_6912664}.

\subsection{Media Bias Fact Check (MBFC)}

To answer our research questions, we not only need the political polarization of a news media outlet, but also an estimate of its factual reliability. In order to get the reliability data we use Media Bias Fact Check, which offers the largest set of labels of any news source rating service~\citep{bozarth2020higher}. For each news media outlet, a minimum of 10 headlines are reviewed and a minimum of 5 news stories are reviewed to get a reasonable factual rating~\footnote{\url{https://mediabiasfactcheck.com/methodology}.}. MBFC classifies reliability in 6 levels: VERY HIGH (a score of 0), which means that the source is considered to be always factual; HIGH (a score of 1 to 2), which means that the source is considered to be almost always factual; MOSTLY FACTUAL (a score of 3 to 4), which means that the source is considered to be usually factual but may have failed a fact check or two that was not promptly corrected; MIXED (a score of 5 to 6), which means the source does not always use proper sourcing or sources to other mixed factual sources; LOW (a score of 7 to 9), which means the source rarely uses credible sources and is not trustworthy for reliable information; VERY LOW (a score of 10), which means the source rarely uses credible sources and is not trustworthy for reliable information. For example, in MBFC the New York Times is rated HIGH as they are mostly reliable except for some Op-Eds and Fox News is rated MIXED because they may publish misleading reports.

\begin{figure}[H]
\centering
\includegraphics[width=0.6\textwidth]{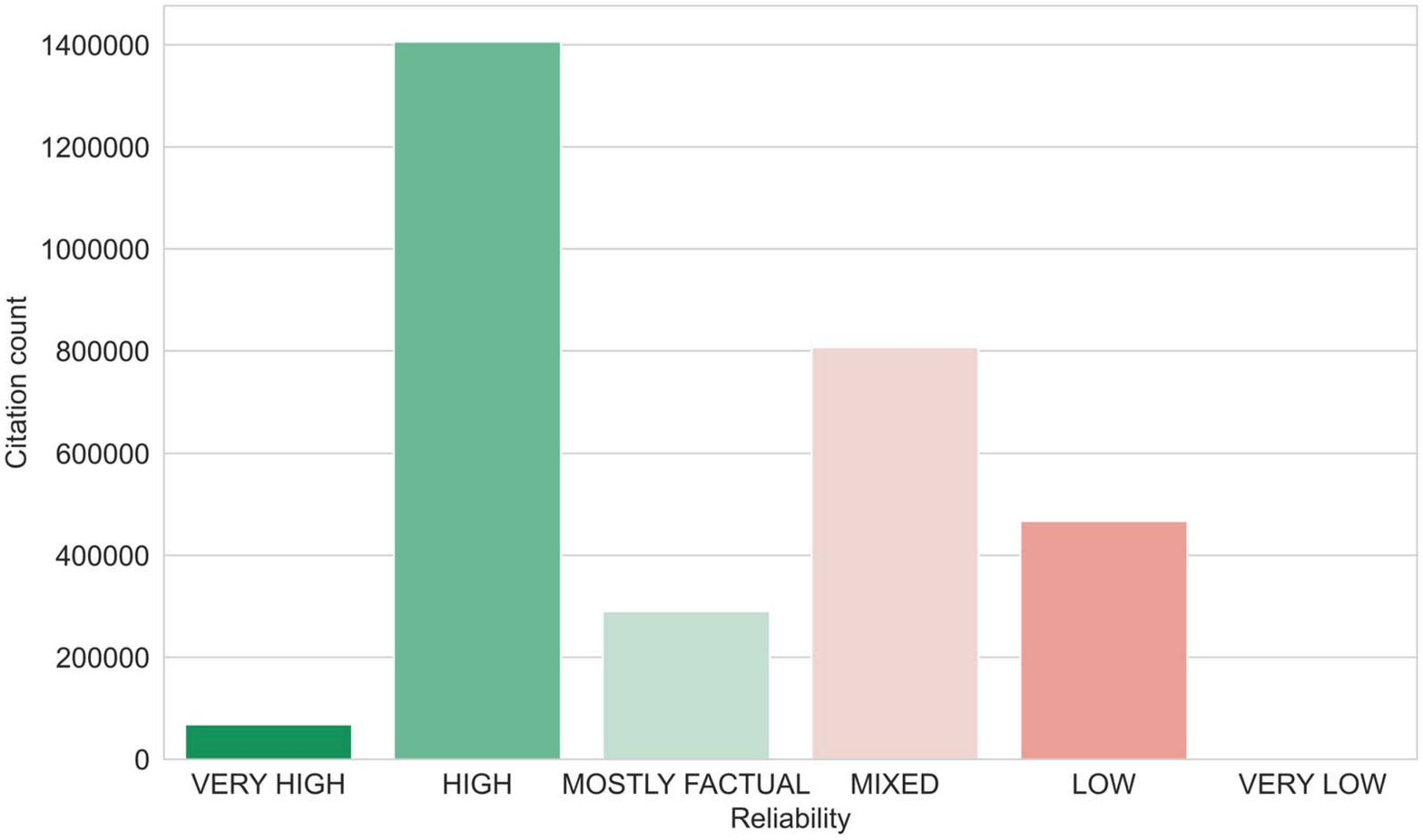}
\caption{Distribution of Wikipedia's news media citation reliability scores.}
\label{fig:Distribution of reliability of citations}
\end{figure}

We crawled all the news media outlet data on MBFC and got a dataset including the ratings for 3,586 outlets. Since we already have the domain names of each URLs in \textit{Wikipedia Citations}, we use the same method to extract the domain names from the MBFC dataset as well. We then match the two datasets via domain names. 689 (19.2\% out of 3,586) domains are matched resulting in 3,041,283 Wikipedia citations with both a factual rating and political polarization score, or 10.4\% of all citations. In Figure~\ref{fig:Distribution of reliability of citations} we show a bar plot of the number of Wikipedia citations by reliability scores, noting that, while there are only 1467 citations rated as ``VERY LOW'', there remains a sizable fraction of citations to low or mixed reliability outlets.

\section{Results}

\begin{figure}[H]
\centering
\includegraphics[width=0.6\textwidth]{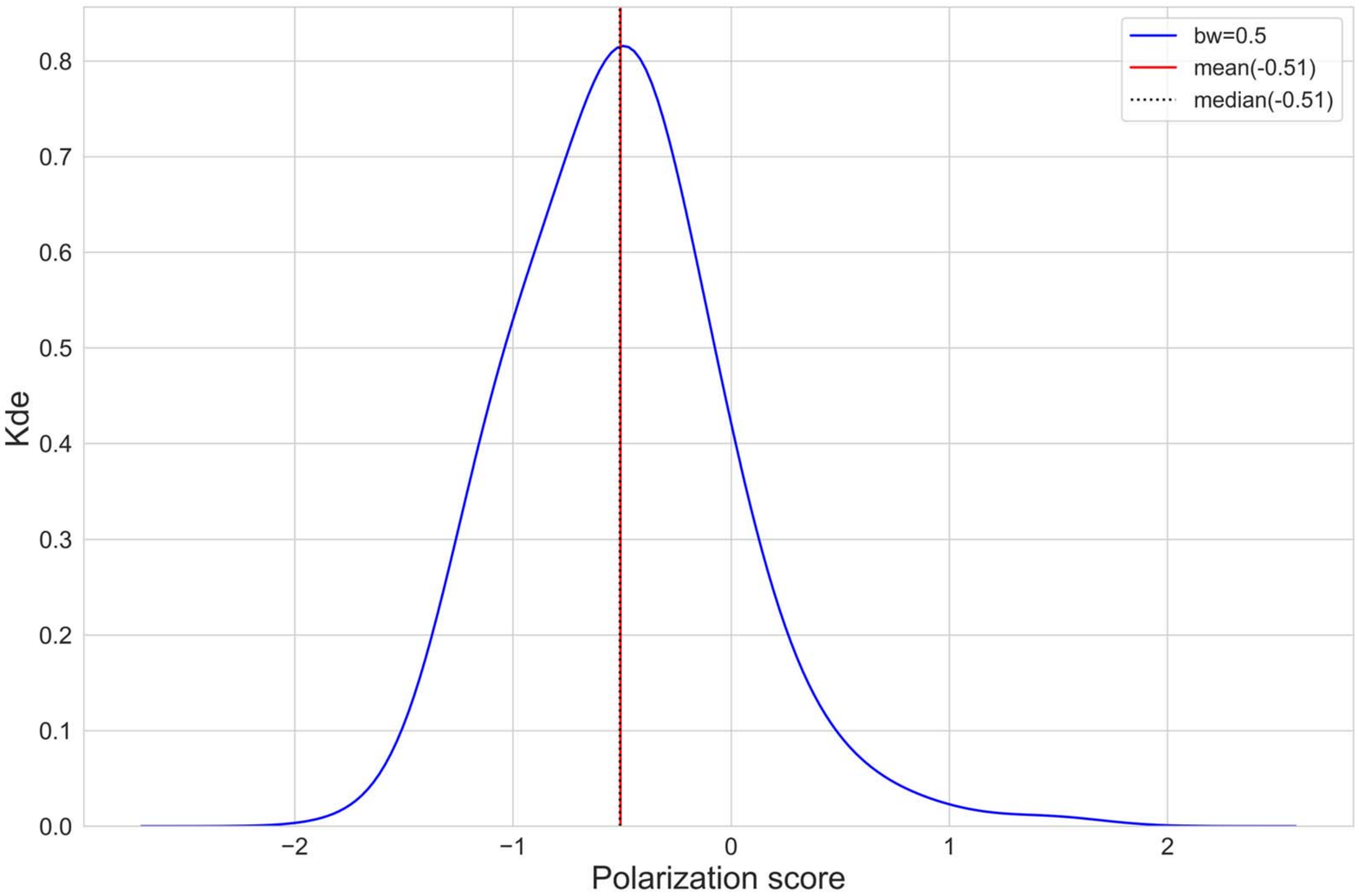}
\caption{Distribution of Wikipedia's news media citation political polarization scores using Kernel Density Estimates (KDE). Negative: liberal; positive: conservative.}
\label{fig:Distribution of bias score and citations}
\end{figure}

We start by providing the overall distribution of Wikipedia's citation political polarization score in Figure~\ref{fig:Distribution of bias score and citations}. We remind that the polarization score (x-axis) ranges between -2 (very liberal) and 2 (very conservative). The average Wikipedia citation polarization score (red line) is -0.51 (median -0.52), therefore leaning towards liberal. The bulk of citations also falls between the range -1 and 0.

\begin{figure}[H]
\centering
\includegraphics[width=0.8\textwidth]{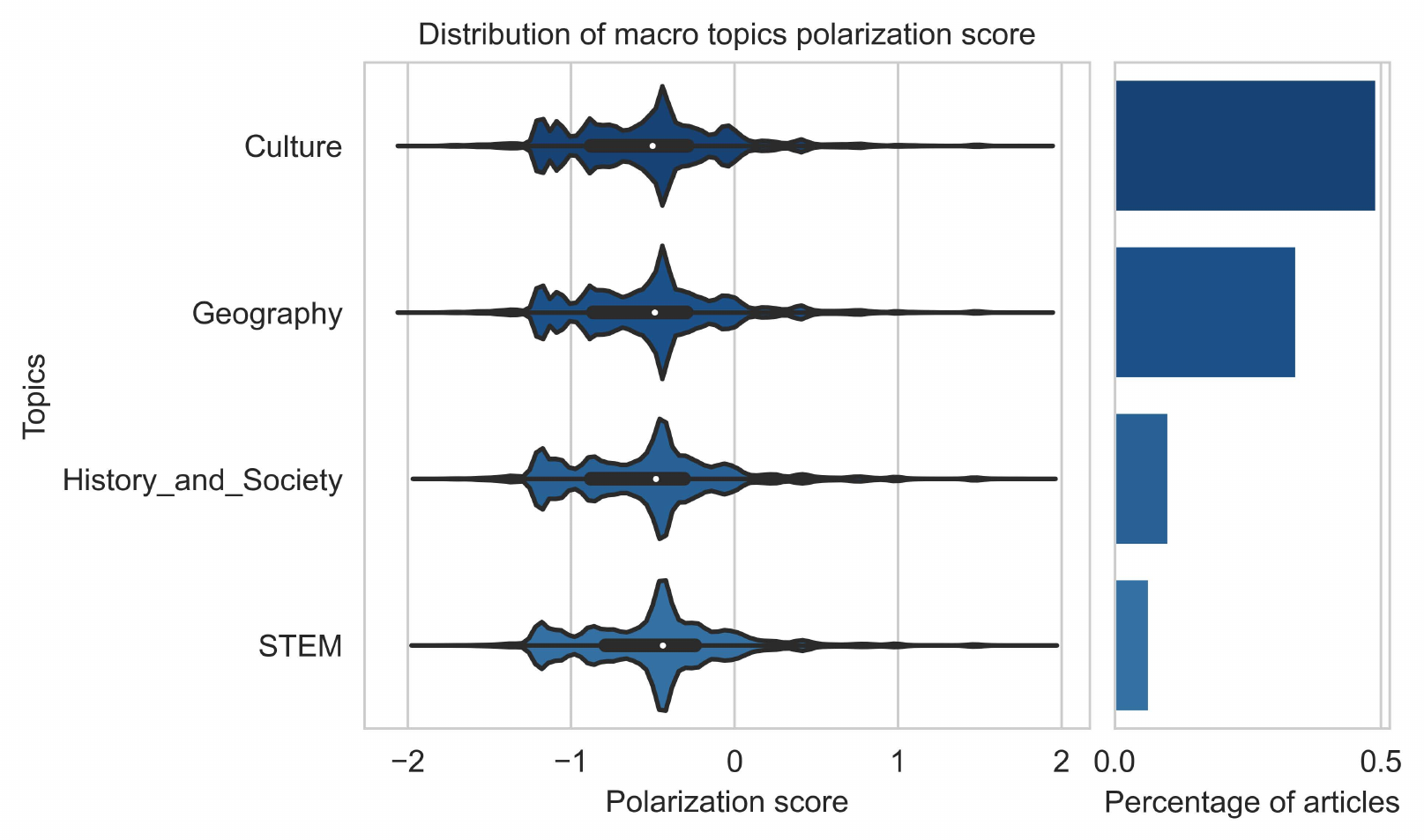}
\caption{Distribution of Wikipedia citation political polarization scores per macro topic. Negative: liberal; positive: conservative.}
\label{fig:Distribution of macro topics bias score}
\end{figure}

We attempt to break down these results using information on Wikipedia's articles, namely their topics and WikiProject. Topics are organised hierarchically, with four macro topics: Culture, Geography, History and Society, STEM. The overview of citation political polarization per macro topic is given in Figure~\ref{fig:Distribution of macro topics bias score}. On the left side, we use a violin plot to show the distribution of polarization scores for each macro topic. From this plot, we cannot see differences among macro topics. The bar plot on the right side provides the relative size of a topic in Wikipedia, showing how articles in Culture takes up nearly 50\% of all citations, while STEM covers 6.5\% of them.

\begin{figure}[H]
\centering
\includegraphics[width=0.8\textwidth]{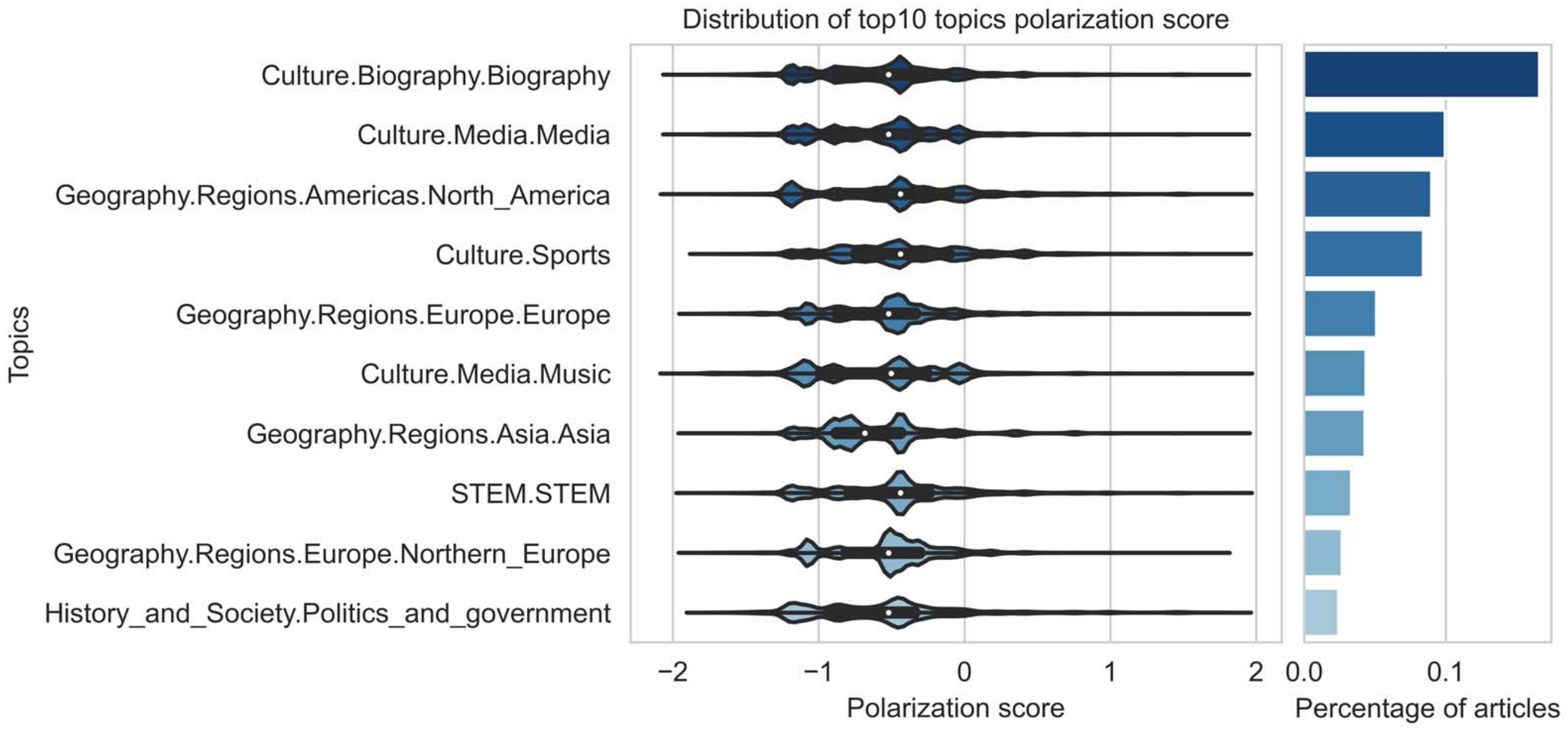}
\caption{Distribution of Wikipedia citation political polarization scores for the top 10 topics. Negative: liberal; positive: conservative.}
\label{fig:Distribution of top10 topics bias score}
\end{figure}

\begin{figure}[H]
\centering
\includegraphics[width=0.8\textwidth]{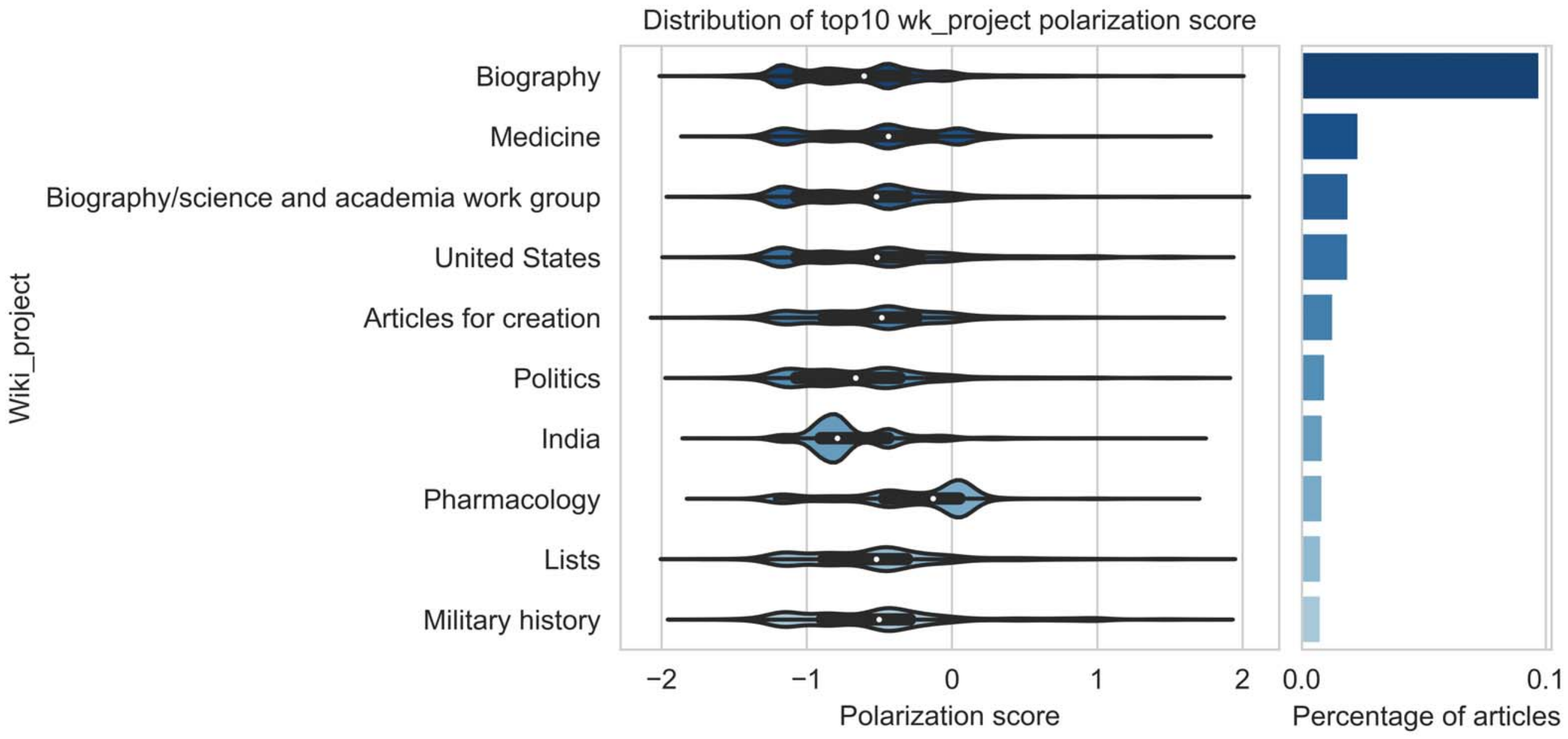}
\caption{Distribution of Wikipedia citation political polarization scores for the top 10 WikiProjects. Negative: liberal; positive: conservative.}
\label{fig:Distribution of top10 wiki_Project bias score}
\end{figure}

Similarly, we show the distribution for the top 10 topics in Figure~\ref{fig:Distribution of top10 topics bias score} and for the top 10 WikiProjects in Figure~\ref{fig:Distribution of top10 wiki_Project bias score}. We again confirm the general trend discussed above, while also finding minor shifts from it. For example, the topic sports has a higher conservative-leaning fraction of citations, all the while maintaining a liberal-leaning skew. The WikiProjects Politics and India are more liberal-leaning than the average, instead. Taken together, these results confirm that the overall trend towards liberal political polarization is not specific to some areas of Wikipedia, but seems to be widespread across topics and WikiProjects.

\begin{figure}[H]
\centering
\includegraphics[width=0.8\textwidth]{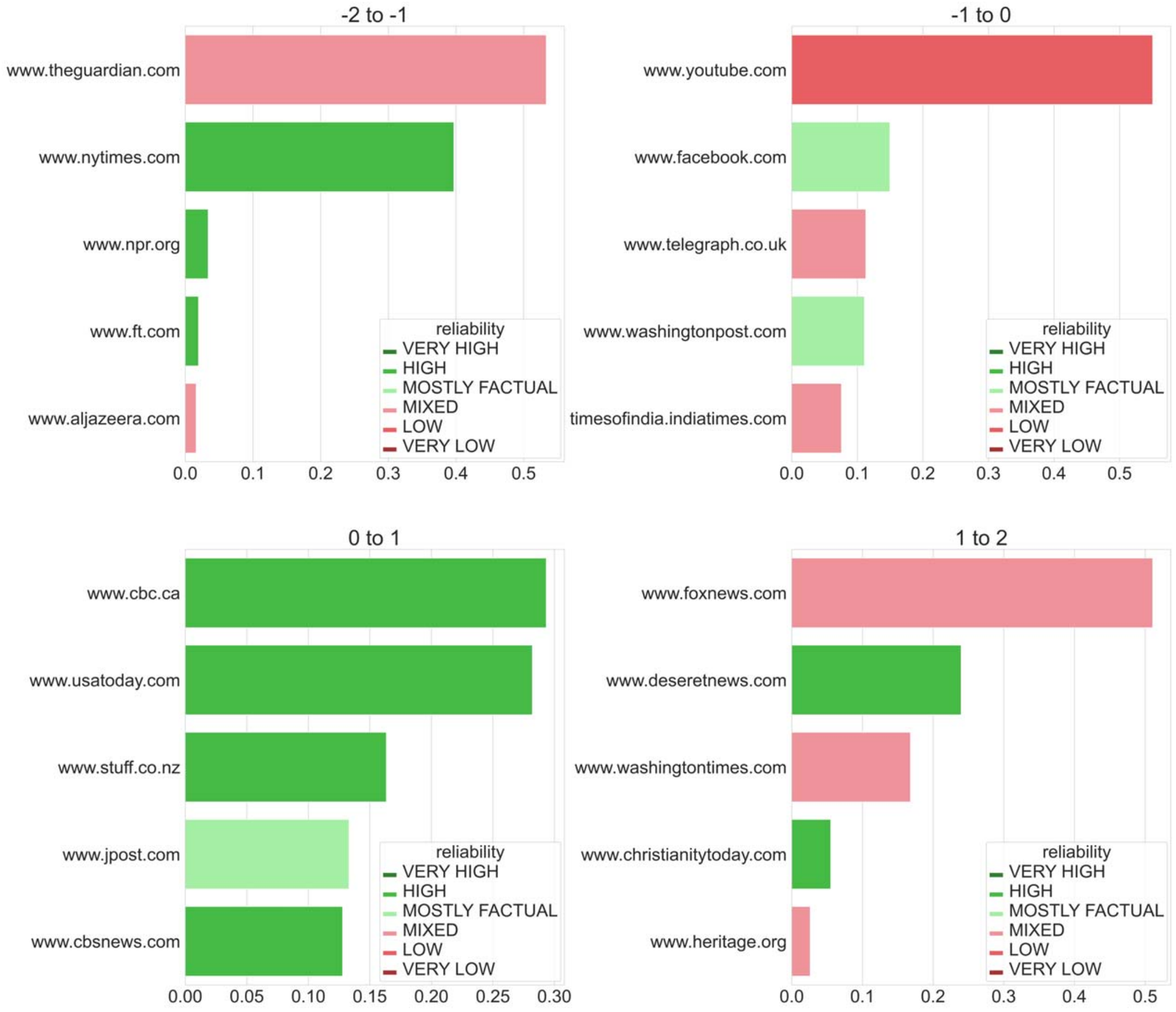}
\caption{Top 5 news outlets reliability class for different political polarization groups.}
\label{fig:Top5 news sources’ reliability in different bias level.}
\end{figure}

\begin{figure}[H]
\centering
\includegraphics[width=0.6\textwidth]{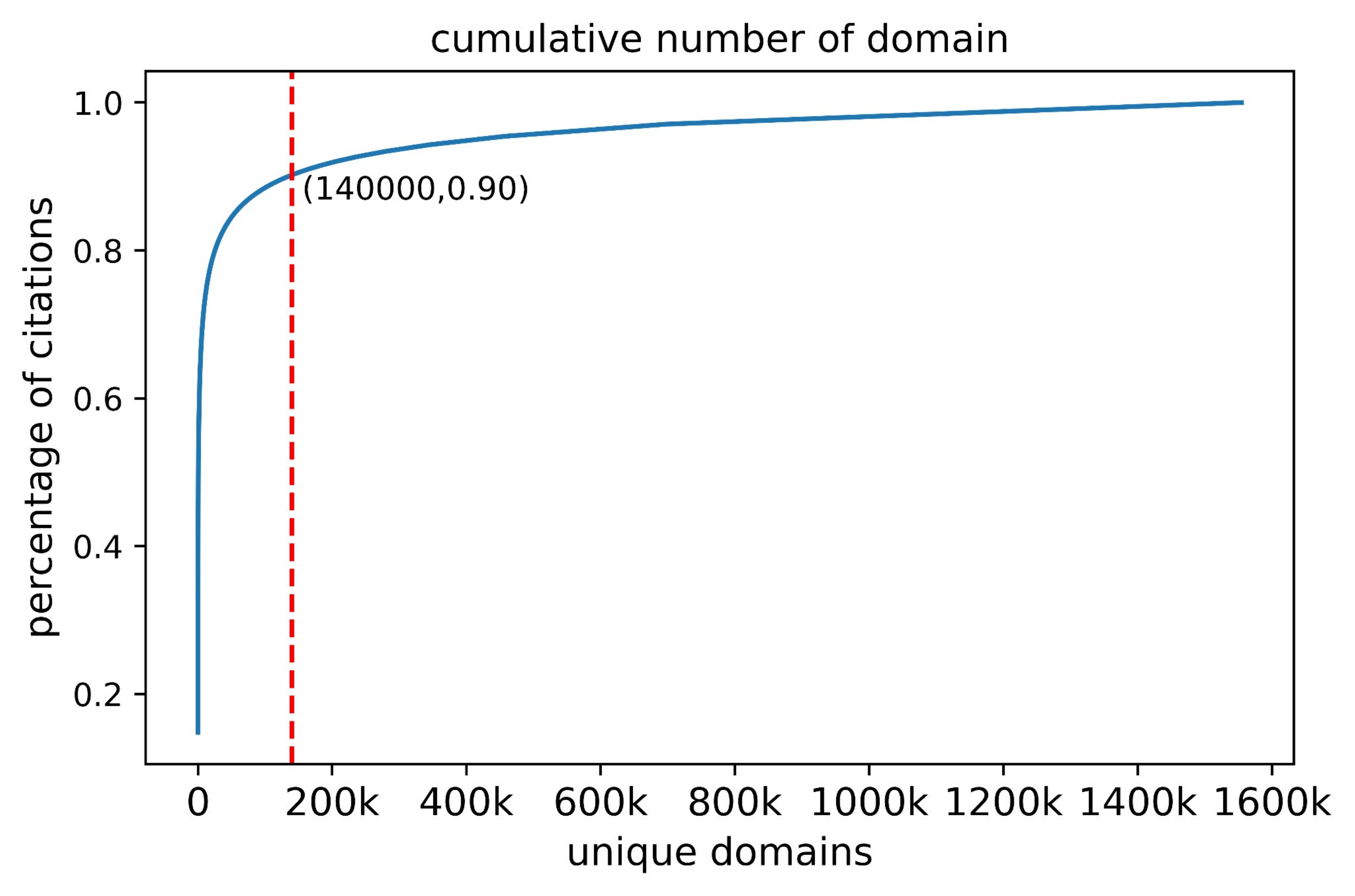}
\caption{Cumulative distribution of Wikipedia citations to distinct domain names.}
\label{fig:cumulative distribution of unique domains}
\end{figure}

\begin{figure}[H]
\centering
\includegraphics[width=0.8\textwidth]{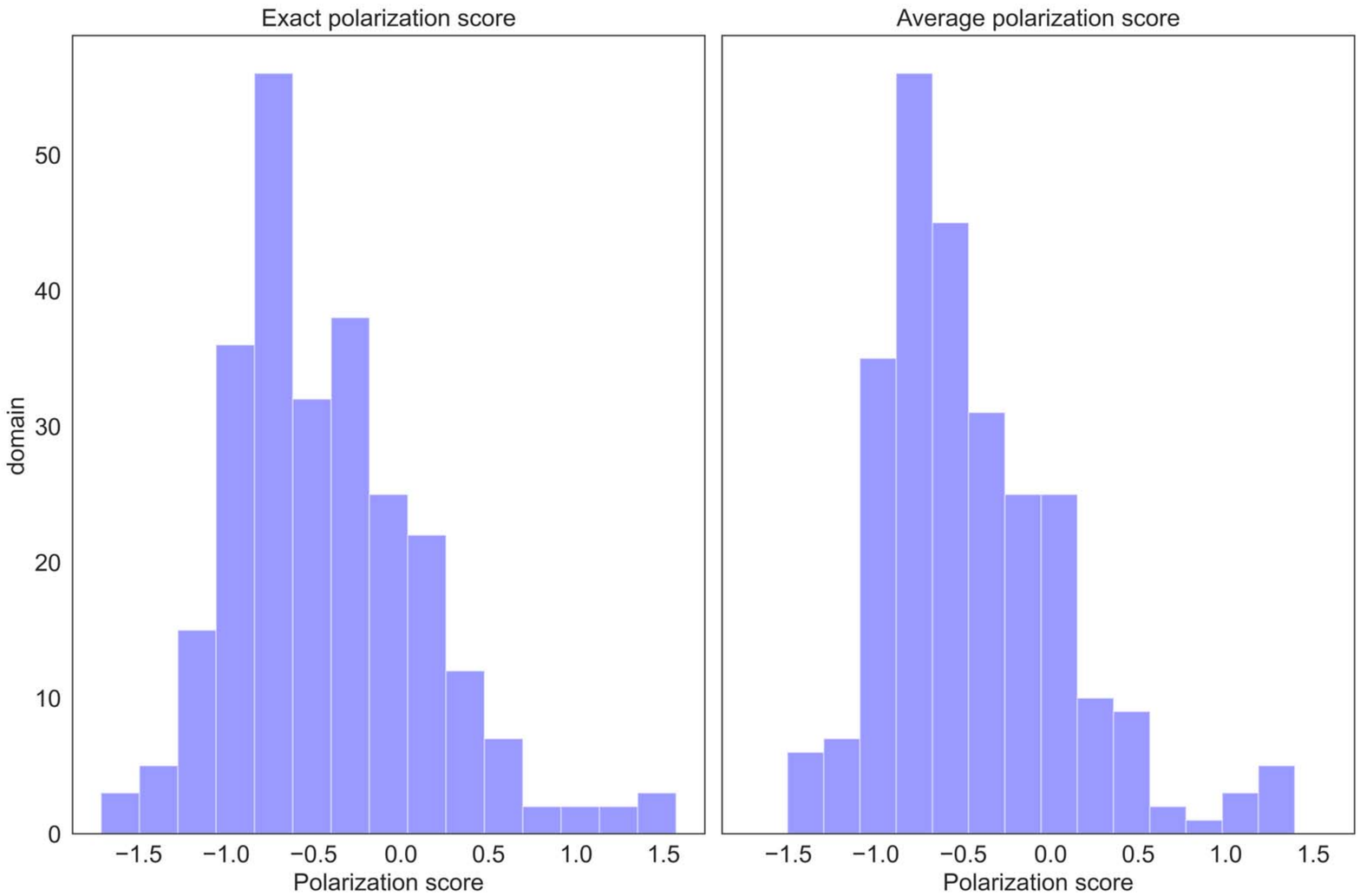}
\caption{Distribution of polarization score for multiple results.}
\label{fig:Distribution of bias score for multiple results}
\end{figure}

\begin{figure}[H]
\begin{subfigure}{.5\textwidth}
  \centering
  \includegraphics[width=.99\linewidth]{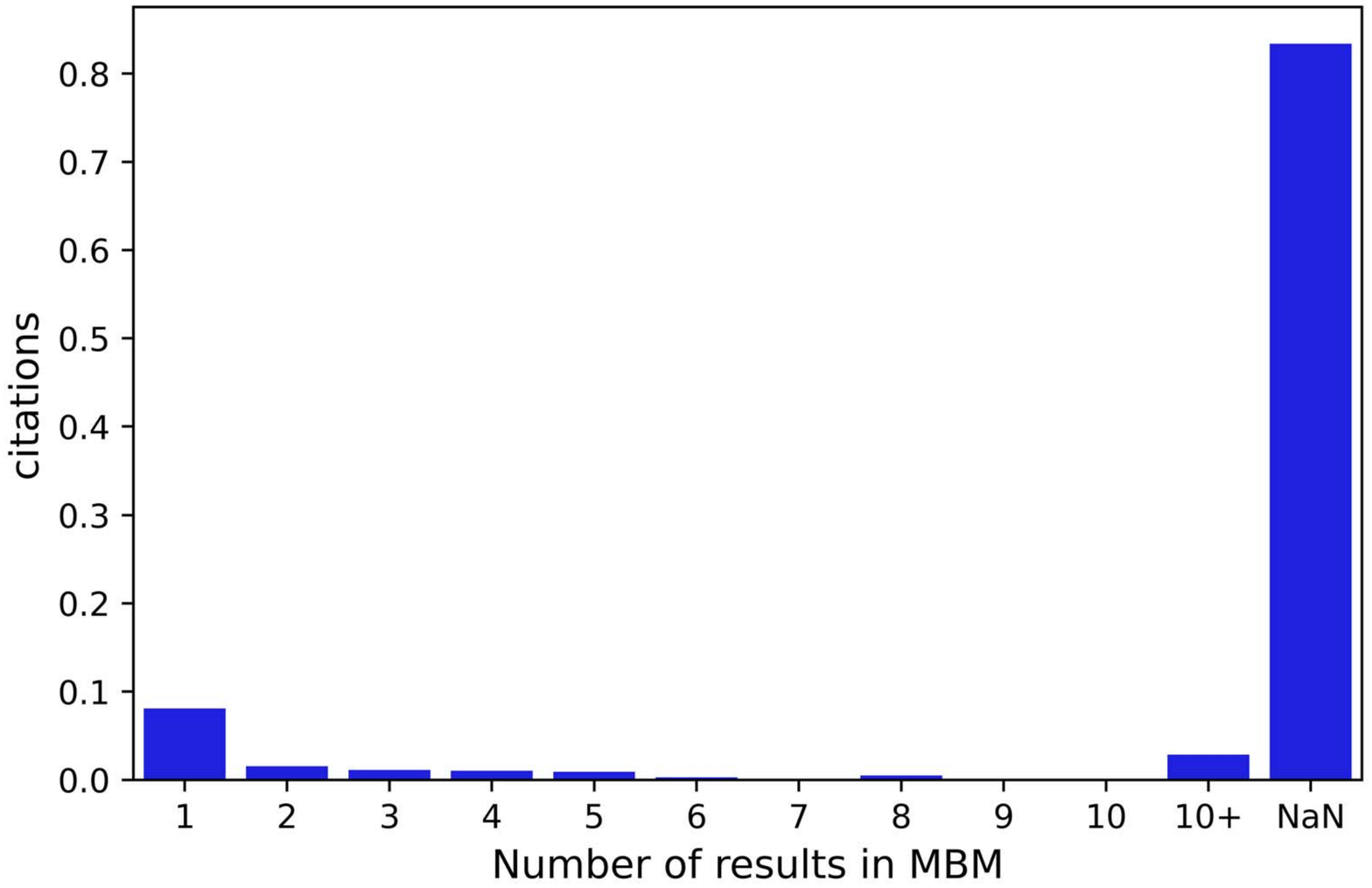}
  \caption{With \textit{NaN}.}
  \label{fig:Figure2_left}
\end{subfigure}%
\begin{subfigure}{.5\textwidth}
  \centering
  \includegraphics[width=.99\linewidth]{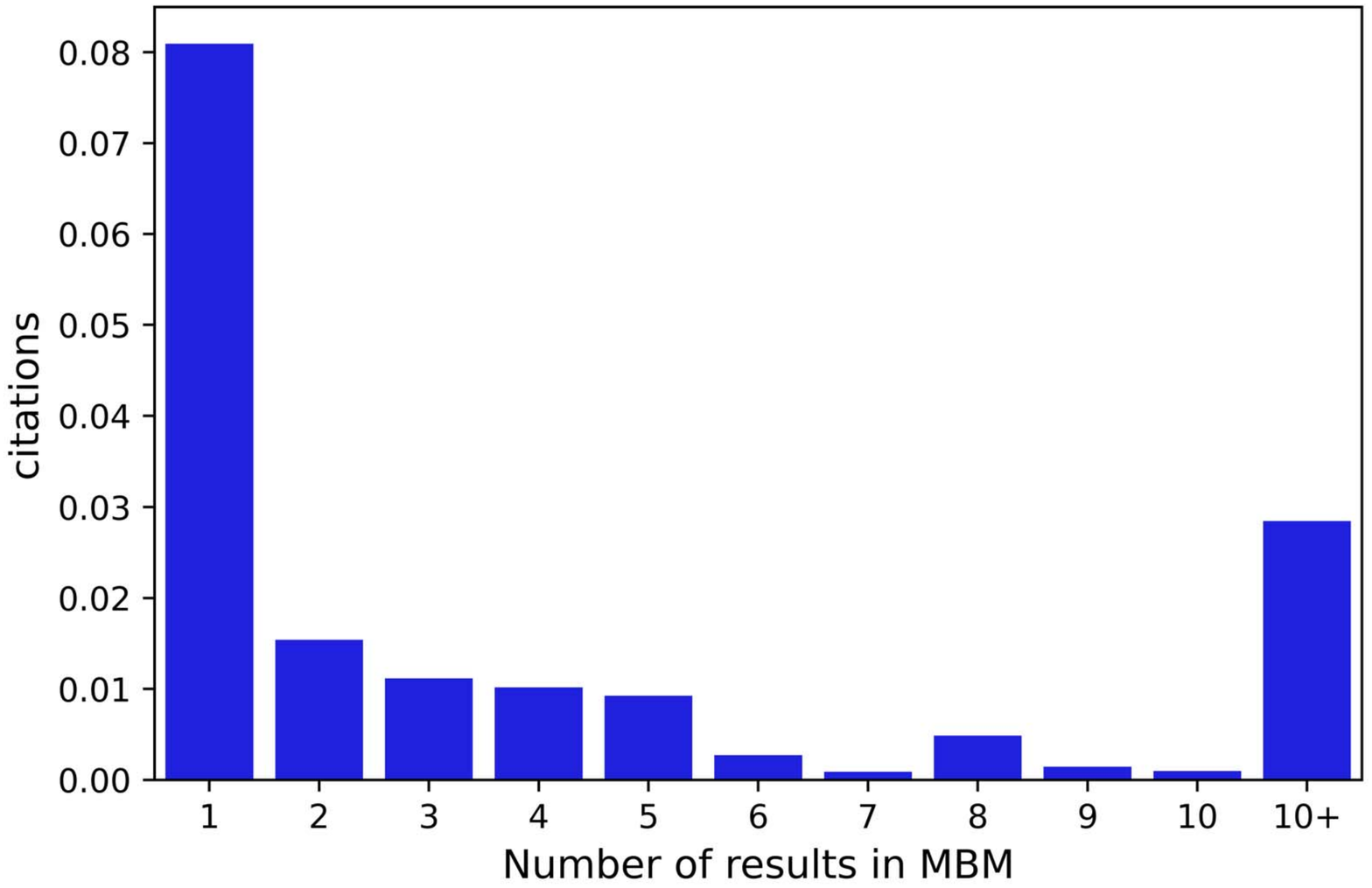}
  \caption{Without \textit{NaN}.}
  \label{fig:Figure2_right}
\end{subfigure}
\caption{Fraction of matches in MBM for a given domain name, and their citation coverage.}
\label{fig:Figure2}
\end{figure}

In principle, in Wikipedia the neutrality and reliability of contents are tied together. Nevertheless, in practice, we speculate that editors may introduce political polarization in their sources in order to prioritise reliable ones. We have shown before, in Figure~\ref{fig:Distribution of reliability of citations}, that most cited news outlets are labelled as highly reliable or mostly factual, even if a significant share of mixed or low reliability sources remains. More details are given in Figure~\ref{fig:Top5 news sources’ reliability in different bias level.}, where we plot the top 5 news outlets per polarization score group, and show their reliability class as well. In this plot, we divide the news outlets into four groups according to polarization scores: Very Liberal [-2 to -1], Liberal (-1 to 0], Conservative (0 to 1], Very Conservative (1 to 2]. For each group, the x-axis shows the percentage of a news outlet by citations within its group. We can see that, for example, The Guardian is labelled as mixed reliability and takes more than 50\% of citations in the Very Liberal group, while the NYT is the second Very Liberal source and is considered highly reliable. Fox News is the top Very Conservative news outlet, with mixed reliability score. We note that in the group Liberal we also have YouTube, with a low reliability score. YouTube is not an outlet with an editorial policy per se, but a repository of contents of any kind. We therefore test whether our results hold when removing citations to YouTube from the dataset, finding that after removal the political polarization distribution moves slightly further liberal overall, while the effect of a conservative polarization in low reliability sources fades substantially (see below). Nevertheless, these changes do not alter out main findings.

Next, we make use of multiple linear regression to address RQ2 and explore whether there is a relationship between political polarization and reliability of news media sources in Wikipedia. In our model, we take the political polarization score as the dependent variable, using Wikipedia article topics and WikiProjects as independent variables and set reliability as a control variable. To simplify the model, we proceed as follows. For topics, we use the macro topic: Geography, History and Society, Culture and STEM. For WikiProjects, we focus on the top 10 WikiProjects and set the rest of WikiProjects as ``Other''. For reliability, we take MOSTLY FACTUAL as the reference class, which is the one in between high and low reliability classes. Thus, our final model is based on the following formula:

\begin{equation*}
\begin{split}
Polarization\ score = & Reliability+\\& Geography + Culture + History\ and\ Society + STEM +\\ & Biography + Medicine + Biography\ science\ and\ academia work\ group +\\ & United\ States + Articles\ for\ creation + Politics + India +\\ & Pharmacology + Lists + Military\ history + Other
\end{split}
\end{equation*}

We provide the results of our regression analysis in Table~\ref{tab:Regression result}. Firstly, while most effects appear to be significant, their coefficient magnitude is always small. We first notice how small the effect of topics is, confirming our previous intuition. Some WikiProjects show slightly larger effects, for example India (more liberal) and Pharmacology (less liberal). Considering reliability in turn, we cannot see a clear pattern emerge. While high reliability shows a liberal skew, very high reliability shows a conservative skew in turn. Mixed sources tend to be more liberal, while low and very low reliability ones tend to be more conservative.

To test our results, we develop several different models. First of all, we test only for polarization score and Wikipedia topics. In this model, all macro topics have a significant effect on the polarization score with small coefficients, that is, Geography and STEM will bring a less liberal skew while Culture History and Society will have an ever stronger liberal skew. When using a model with reliability and topics, our results converge and become very similar to the model discussed above which also includes WikiProjects. As mentioned previously, we also test our final model without citations to YouTube. After removing them, the most important change is that the low reliability coefficient becomes non significant and goes close to zero, thus making the case for a possible association between low reliability and conservative news outlets disappear. 

\begin{table}[H]
\caption{Regression results for the effect of news media reliability on political leaning, controlling for Wikipedia topics and projects.}\label{tab:Regression result}
\resizebox{\textwidth}{!}{%
\begin{tabular}{l|llllll}
Variables                                                          & coef  & std err & t        & P\textgreater{}|t| & {[}0.025 & 0.975{]} \\ \hline
Intercept                                                          & -0.56 & 0.005   & -118.360 & 0.000              & -0.569   & -0.550   \\
C(factual, Treatment(reference='MOSTLY FACTUAL')){[}T.HIGH{]}      & -0.09 & 0.002   & -37.653  & 0.000              & -0.090   & -0.081   \\
C(factual, Treatment(reference='MOSTLY FACTUAL')){[}T.LOW{]}       & 0.09  & 0.003   & 30.143   & 0.000              & 0.083    & 0.095    \\
C(factual, Treatment(reference='MOSTLY FACTUAL')){[}T.MIXED{]}     & -0.20 & 0.002   & -82.082  & 0.000              & -0.201   & -0.192   \\
C(factual, Treatment(reference='MOSTLY FACTUAL')){[}T.VERY HIGH{]} & 0.16  & 0.004   & 42.224   & 0.000              & 0.149    & 0.163    \\
C(factual, Treatment(reference='MOSTLY FACTUAL')){[}T.VERY LOW{]}  & 0.07  & 0.023   & 2.833    & 0.005              & 0.020    & 0.112    \\
Geography                                                          & 0.03  & 0.002   & 16.010   & 0.000              & 0.023    & 0.030    \\
Culture                                                            & -0.00 & 0.002   & -0.758   & 0.448              & -0.004   & 0.002    \\
History and Society                                                & -0.02 & 0.001   & -16.579  & 0.000              & -0.027   & -0.021   \\
STEM                                                               & 0.01  & 0.002   & 7.098    & 0.000              & 0.009    & 0.016    \\
Biography                                                          & -0.04 & 0.002   & -23.958  & 0.000              & -0.047   & -0.039   \\
Medicine                                                           & 0.01  & 0.004   & 2.037    & 0.042              & 0.000    & 0.015    \\
Biography\_science\_and\_academic\_work\_group                     & -0.04 & 0.004   & -10.502  & 0.000              & -0.047   & -0.032   \\
United\_States                                                     & 0.06  & 0.002   & 36.096   & 0.000              & 0.060    & 0.067    \\
Articles\_for\_creation                                            & 0.01  & 0.005   & 2.874    & 0.004              & 0.004    & 0.022    \\
Politics                                                           & -0.01 & 0.002   & -6.233   & 0.000              & -0.017   & -0.009   \\
India                                                              & -0.09 & 0.004   & -24.027  & 0.000              & -0.101   & -0.086   \\
Pharmacology                                                       & 0.11  & 0.007   & 14.672   & 0.000              & 0.092    & 0.120    \\
Lists                                                              & 0.04  & 0.003   & 15.955   & 0.000              & 0.037    & 0.047    \\
Military\_history                                                  & 0.03  & 0.002   & 11.020   & 0.000              & 0.022    & 0.032    \\
Other                                                              & 0.02  & 0.004   & 3.986    & 0.000              & 0.008    & 0.023   \\ \hline
\multicolumn{7}{l}{No. Observations: 604459 \ \ \ \ \ \ \ \ \ \ \ \ \ \ \ \ \ \ \ \ \ \ \ \ \ \ \ \ \ \ \ \ \ \ \ \ \ \ \ \ \ \ \ \ \ \ \ \ \ \ \ \ \ \ \ \ \ \ \ \ \ \ \ \ \ \ \ \ \ \ \ \ \ \ \ \ \ \ R-squared: 0.047}
\end{tabular}%
}
\end{table}

\section{Discussion}

Wikipedia editors follow core policies when editing articles~\citep{pavalanathan2018mind}, in an attempt to provide a neutral point of view and reliable contents~\citep{mesgari2015sum}. Nevertheless, biases might still be found in Wikipedia in a variety of forms, and as such they require a never-ending effort on the part of the community. We find a moderate yet systematic liberal polarization in Wikipedia's news media sources. The average polarization score of Wikipedia sources is -0.5, with the distribution of polarization scores concentrated around -1 to 0, on a scale between -2 (very liberal) and 2 (very conservative). Our results partially confirm and extend previous ones~\citep{Agarwal}, while also showing that Wikipedia remains polarized towards liberal news media~\citep{gentzkow2010drives}. This finding is relevant as it signals a possible systematic biasing effect whose causes and effects will have to be further studied. News media sources not only select and provide specific information, but also convey it with a certain framing which might influence how a topic is discussed in Wikipedia.

We initially speculated that the presence of political polarization might be partially explained by the editors' need to balance a source factual reliability with its political leaning. Interestingly, we find no clear relationship between reliability and polarization. The relationship between reliability and political polarization is complex, with more conservative sources being associated with both high and low reliability, while liberal sources tend to more often be of mixed reliability. This finding leaves the question of the motivations for political polarization open. We speculate that a multiplicity of factors might play a role, from pre-existing leanings in the composition of the editors' community, to an increasingly polarized media landscape making it difficult to find neutral news media sources to use. Our results may also help the case for changing Wikipedia's sourcing policies, which might prevent information lacking accepted reliable secondary sources from being considered, and at the same time prevent broad and important areas of knowledge from entering the Wikipedia project. This is especially the case for cultures that rely on the non-written transmission and expression of knowledge (such as oral sources)~\citep{wikipedia_Knowledge_Integrity}. Relying on a more diverse set of sources could be an approach to reduce possible bias in Wikipedia.

We acknowledge several limitations of our study, some of which constitute possible directions for future work. First of all, we rely also on external sources to measure political polarization and reliability. While such sources are considered authoritative, their coverage is only partial and they use specific approaches to score news media outlets which could be complemented in the future. We also mainly focus on the binary political polarization distinction between liberal and conservative sources. Several other political dimensions exist which could be considered in the future. Exploring the relationship between source polarization and reliability remains an open challenge, one that should focus on more dimensions than what we considered here, including time. In this respect, our work scores news media sources at the domain level, while a more granular analysis should be done at the level of the individual source contents. Similarly, we did not consider how a source is used in Wikipedia, and whether its use reflects such polarization or not. Our study remains focused on English Wikipedia, while its extension to more languages would provide for a broader picture.

\section{Conclusion}

In this contribution, we analyzed a potential source of bias in Wikipedia, by considering citations to news media sources and their political polarization. We used a large-scale dataset of citations from Wikipedia, enriching it with metrics of political media polarization from the Media Bias Monitor, and of factual reliability from the Media Bias Fact Check. We found a moderate yet systematic liberal polarization in Wikipedia's news media sources. We also showed that there is no clear relationship between a news media source's reliability and its political leanings.

These results offer a foundation to inform Wikimedia's research agenda about possible sources of disinformation and bias, in view of upholding its neutral point of view policy. Specifically, to keep Wikipedia as a neutral source of information, a better understanding of possible sources of bias is needed not only considering contents or editors, but also Wikipedia's external sources. Here we provided a preliminary analysis of political polarization in Wikipedia from the perspective of citations to news media sources, while much works remain to be done. On the one hand, the measurement of political polarization, reliability and other potential signals of bias could be considered more comprehensively, although we have relied here on authoritative sources. On the other hand, a more granular study on the level of the contents of sources and how they are used in Wikipedia would significantly enrich our preliminary findings. The relationship between polarization and reliability would benefit from a higher dimensional analysis. Therefore, we see our work as fostering further attention to the always-open challenge of preserving and improving Wikipedia's knowledge integrity.

\subsection{Code and data availability statement}

The code to replicate our work is available online: \url{https://github.com/alsowbdxa/News_Bias_and_reliability_in_Wikipedia}. The Wikipedia Citations dataset is also openly available~\citep{singh2021wikipedia}. All other supporting datasets we used are available and referenced from the Data and Methods section.  


\section{Appendix}

\bibliographystyle{unsrt}


\end{document}